\documentclass[aps,reprint,amsmath,amssymb,superscriptaddress,showpacs,prl]{revtex4-1}
\usepackage{graphicx}
\usepackage{color}
\usepackage{lipsum}
\usepackage{dcolumn}
\begin{document}

\title{Comment on ``Interplay of Structural and Optoelectronic Properties in
Formamidinium Mixed Tin–Lead Triiodide Perovskites''}

\author {S.~D.~Baranovskii}
\affiliation{Faculty of Physics and Material Sciences Center,
Philipps Universit\"{a}t, D-35032 Marburg, Germany}
\affiliation{Department f\"{u}r Chemie, Universit\"{a}t zu K\"{o}ln, Luxemburger Strasse 116, 50939 K\"{o}ln, Germany}

\author {P.~H\"{o}hbusch}  %
\thanks{Trainee at the Faculty of Physics, Philipps-University, D-35032 Marburg, Germany}
\affiliation{Faculty of Physics and Material Sciences Center,
Philipps Universit\"{a}t, D-35032 Marburg, Germany}

\author{A.~V.~Nenashev}
\affiliation{Faculty of Physics and Material Sciences Center,
Philipps Universit\"{a}t, D-35032 Marburg, Germany}
\affiliation{Rzhanov Institute of Semiconductor Physics, 630090 Novosibirsk, Russia}
\affiliation{Novosibirsk State University, 630090 Novosibirsk, Russia}

\author{A.~V.~Dvurechenskii}
\affiliation{Rzhanov Institute of Semiconductor Physics, 630090 Novosibirsk, Russia}
\affiliation{Novosibirsk State University, 630090 Novosibirsk, Russia}

\author{M.~Gerhard}
\affiliation{Faculty of Physics and Material Sciences Center,
Philipps Universit\"{a}t, D-35032 Marburg, Germany}

\author{M.~Koch}
\affiliation{Faculty of Physics and Material Sciences Center,
Philipps Universit\"{a}t, D-35032 Marburg, Germany}

\author{D.~Hertel}
\affiliation{Department f\"{u}r Chemie, Universit\"{a}t zu K\"{o}ln, Luxemburger Strasse 116, 50939 K\"{o}ln, Germany}

\author{K.~Meerholz}
\affiliation{Department f\"{u}r Chemie, Universit\"{a}t zu K\"{o}ln, Luxemburger Strasse 116, 50939 K\"{o}ln, Germany}

\author{F.~Gebhard}
\affiliation{Faculty of Physics and Material Sciences Center,
Philipps Universit\"{a}t, D-35032 Marburg, Germany}

\date{\today}

\date{\today}

\begin{abstract}
Studying optoelectronic properties in FAPb$_{1-x}$Sn$_x$I$_3$ perovskites as a function of the lead:tin content,
  Parrott et al.\ observed the broadest luminescence linewidth and the largest luminescence Stokes shift
  in mixed compositions with Sn $ < 25$\% and
  with $> 0.85$\%. Since the largest effects of alloy disorder were expected for
  the 50:50 composition, it was concluded that the revealed disorder effects
  might arise from extrinsic factors that can be eliminated upon
  further crystal growth optimization. This comment shows that the largest effects of alloy disorder for perfectly random fluctuations
  in FAPb$_{1-x}$Sn$_x$I$_3$ perovskite are, in fact, expected
  for $x < 0.25$ and for $x > 0.85$.
  Therefore, further crystal growth optimization is futile.

  \end{abstract}

\maketitle   

Alloyed lead-tin perovskites exhibit lower bandgaps than the compositions
with only lead, or only tin, metal cations.
This feature makes alloys promising for applications in tandem solar cells to absorb the low-energy portions
of the solar spectrum. The price for the tunable bandgap is a disorder potential caused by alloying.
In this journal~\cite{Parrott2018},
  Parrott et al.\ published results
on a series of FAPb$_{1-x}$Sn$_x$I$_3$ alloys showing that the impact
of disorder is maximal around mole fractions (alloy compositions) $x_a < 0.25$ and $x_b > 0.85$,
as evidenced by the contribution of the disorder potential
to the luminescence linewidth $\Delta \varepsilon$
and to the luminescence Stokes shift $\langle\varepsilon\rangle_{\mathrm{Stokes}}$ (see Fig.\ 3a,b
in Ref.~\cite{Parrott2018}). This observation was in striking contrast to
the expected in Ref.~\cite{Parrott2018} maximal influence of the disorder
at $x=0.5$. The discrepancy between observations and
expectations was attributed to extrinsic sources of disorder, which might be improved
upon further crystal growth optimization~\cite{Parrott2018}.
In this comment we show that the observations of maximal disorder
effects around $x_{a,b}$
are to be expected for perfectly random potential fluctuations.
Thus, crystal growth optimization will not change
the observed disorder effects.

Statistical fluctuations of the local alloy composition around the average value $x$ cause a disorder potential
acting on electrons and on holes. The band edge of the conduction band, $E_c$,
and that of the valence band, $E_v$, depend on the composition $x$.
Consequently, their derivatives,
$\alpha_{c(v)} = dE_{c(v)}/dx$,
measure the amplitude of the potential fluctuations caused by statistical alloy disorder.
Several theoretical studies~\cite{Alferov1969,Baranovskii1978,Raikh1990,Wiemer2016} have shown
that, for perfectly random fluctuations, the energy scale $\varepsilon_0(x)$ for alloy disorder
in three dimensions is determined by the following combination of
material parameters,
\begin{equation}
\label{eq:scale_disorder}
\varepsilon_0(x) = \frac{[\alpha(x)]^4 x^2(1-x)^2 m^3}{\hbar^6 N^2} \,,
\end{equation}
where $N$~is the concentration of lattice sites occupied by alloy atoms, $m$~is the effective mass of charge carriers,
$\hbar$~is the reduced Planck constant, and $\alpha$ is the derivative of the band edge with respect to $x$. In two dimensions, a slightly different combination
of material parameters determines the scale of energy disorder~\cite{Wiemer2016,Masenda2021}.

Equation~(\ref{eq:scale_disorder}) permits to elucidate the dependence of the
disorder energy scale $\varepsilon_0$ on the alloy composition~$x$
in the lead–tin perovskite FAPb$_{1-x}$Sn$_x$I$_3$. If the parameters $\alpha$, $m$ and $N$
did not depend on~$x$,  $\varepsilon_0(x)$
would solely be proportional to $x^2(1-x)^2$ and would display a
  maximum at $x = 0.5$. However, the parameter $\alpha$ in FAPb$_{1-x}$Sn$_x$I$_3$
  strongly depends on $x$, $\alpha\equiv \alpha(x)$.
  It enters Equation~(\ref{eq:scale_disorder}) in the forth power, which
  drastically alters the composition dependence of the disorder energy scale
    $\varepsilon_0(x)$. In order to calculate $\alpha$, let us use the dependence of the band gap $E_g(x)$
    on $x$, estimating $\alpha$ as
\begin{equation}
  \alpha = dE_g/dx \, .
  \label{eq:alphadef}
\end{equation}
In their comprehensive study of the absorption edge, Parrott et al.~\cite{Parrott2018}
    deduced the dependence $E_g(x)$ in the form
\begin{equation}
\label{eq:bandgap}
E_g(x) = x E^{\rm Sn}_{g} + (1-x) E^{\rm Pb}_{g} - x(1-x) b \,,
\end{equation}
with $b = 0.73\, {\rm eV}$, $E^{\rm Pb}_{g} \approx 1.55\, {\rm eV}$, $E^{\rm Sn}_{g} \approx 1.39\, {\rm eV}$
for the room-temperature phase,
and $b = 0.84$, $E^{\rm Pb}_{g} \approx 1.47\, {\rm eV}$, $E^{\rm Sn}_{g} \approx 1.21\, {\rm eV}$
for the low-temperature phase.

\begin{figure}
\includegraphics[width=\linewidth]{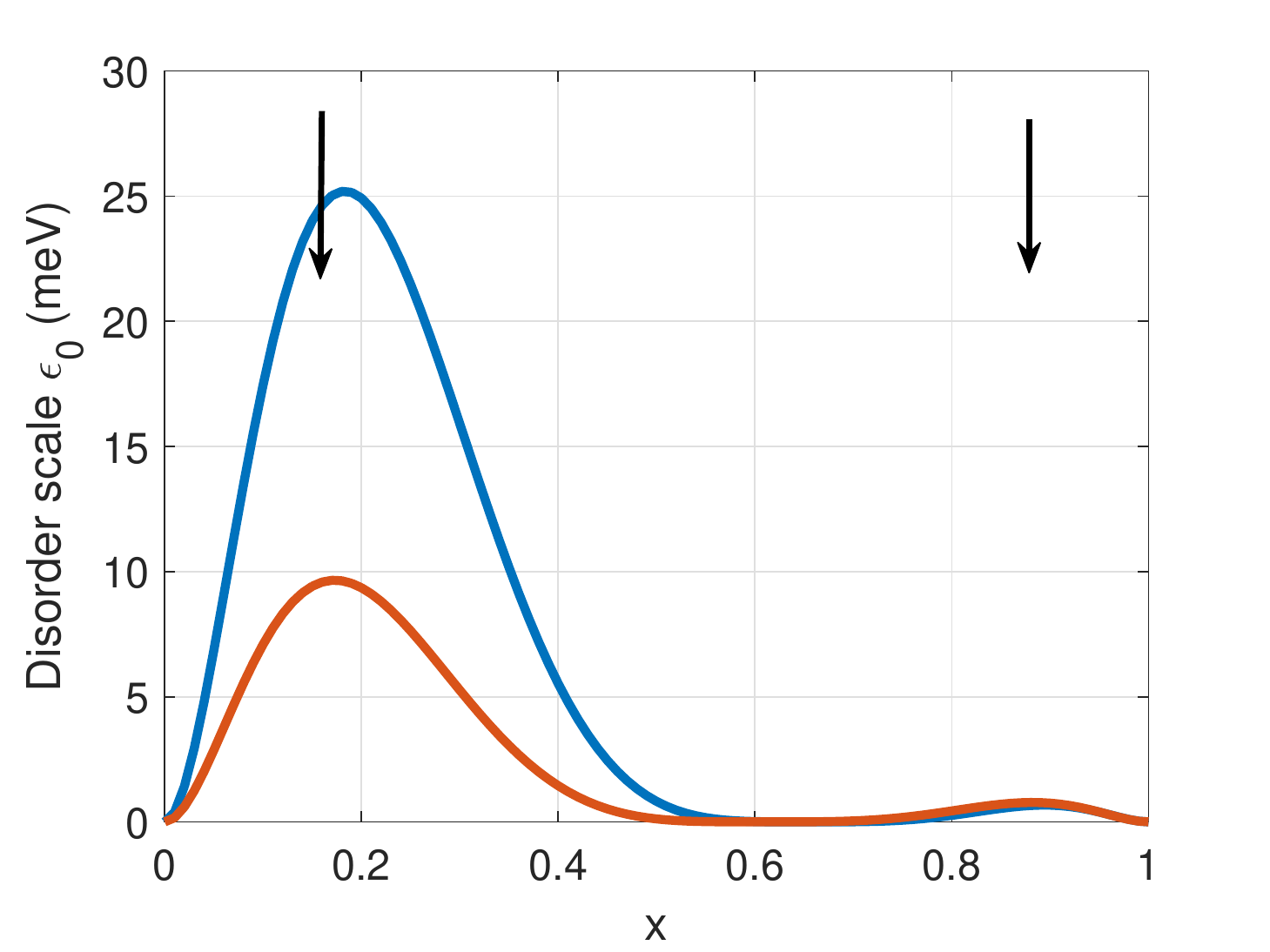}
\caption {Dependence of the disorder energy scale $\varepsilon_0(x)$
    on the composition~$x$, as predicted by Equations (\ref{eq:scale_disorder}), (\ref{eq:alphadef}),
    and (\ref{eq:bandgap}) for the low-temperature phase (blue line) and for the room-temperature phase (red line).
    The arrows point at the compositions that correspond to maximal disorder effects
    revealed in Ref.~\cite{Parrott2018}. }
\label{fig:1}
\end{figure}

Using these data,  Equations (\ref{eq:scale_disorder}), (\ref{eq:alphadef}),
and~(\ref{eq:bandgap}) lead to $\varepsilon_0(x)$ shown in Figure~1
(blue line: low-temperature phase; red line:
room-temperature phase).
In the calculations, a concentration of lattice sites $N = 4 \cdot 10^{21}\, {\rm cm}^{-3}$
was used,
based on the estimates
for the lattice constant $a \simeq 6.3 \cdot 10^{-8}\, {\rm cm}$~\cite{Weber2016,Zong2017}. There is uncertainty in the choice of the effective mass $m$ in Equation~(\ref{eq:scale_disorder}). If electrons and holes appear in the form of excitons, the sum $m_e + m_h$ should be used. In the opposite case, either $m_e$, or $m_h$ should enter the equation. Since $m_e$ and $m_h$ in such perovskites are close to each other ~\cite{Muhammad2020,Wang2020} we use in Equation~(\ref{eq:scale_disorder}), for the sake of certainty, the hole effective mass $m_h = 0.273 \, m_0$.

Figure~1 shows that the energy scale of alloy disorder $\varepsilon_0(x)$
has two maxima at $x_a < 0.25$ and $x_b > 0.85$,
in excellent agreement with the observations of Parrott et al.~\cite{Parrott2018}
for the $x$-dependences of the luminescence linewidth $\Delta \varepsilon(x)$
and of the luminescence Stokes shift $\langle\varepsilon\rangle_{\mathrm{Stokes}}(x)$
as a function of the Sn concentration~$x$.

A fit of  $\Delta \varepsilon(x)$ and of $\langle\varepsilon\rangle_{\mathrm{Stokes}}(x)$, as measured by Parrott et al., requires
quantitative relations between $\varepsilon_0(x)$ on the one hand, and $\Delta \varepsilon(x)$ and
$\langle\varepsilon\rangle_{\mathrm{Stokes}}(x)$ on the other hand.
Computer simulations~\cite{DalDon2004,Wright2017} predict that
$\Delta \varepsilon(x) \approx \beta \varepsilon_0(x)$ and $\langle\varepsilon\rangle_{\mathrm{Stokes}}(x)
\approx \gamma \varepsilon_0(x)$ with factors $\beta$ and $\gamma$ varying
in the range $2 \lesssim \beta,\gamma \lesssim 8$,
depending on material parameters and on temperature.
Without requiring the precise values for~$\beta$ and~$\gamma$,
  this comment elucidates the excellent agreement between the values $x_{a,b}$ for the
maxima of the disorder energy scale $\varepsilon_0(x)$, as predicted by
Equations (\ref{eq:scale_disorder}), (\ref{eq:alphadef}), and (\ref{eq:bandgap}),
and the Sn concentrations
for the maximal disorder effects in the luminescence linewidth
$\Delta \varepsilon(x)$ and in the
Stokes shift $\langle\varepsilon\rangle_{\mathrm{Stokes}}(x)$,
observed by Parrott et al.~\cite{Parrott2018}.
Since the
  theory is based on the assumption of perfectly random potential fluctuations, the nice
  agreement between theory and experiment shows that the crystal growth conditions
  for FAPb$_{1-x}$Sn$_x$I$_3$ \cite{Parrott2018} are already optimal, and
  further crystal growth optimization is futile.


\begin{thebibliography}{12}%
\makeatletter
\providecommand \@ifxundefined [1]{%
 \@ifx{#1\undefined}
}%
\providecommand \@ifnum [1]{%
 \ifnum #1\expandafter \@firstoftwo
 \else \expandafter \@secondoftwo
 \fi
}%
\providecommand \@ifx [1]{%
 \ifx #1\expandafter \@firstoftwo
 \else \expandafter \@secondoftwo
 \fi
}%
\providecommand \natexlab [1]{#1}%
\providecommand \enquote  [1]{``#1''}%
\providecommand \bibnamefont  [1]{#1}%
\providecommand \bibfnamefont [1]{#1}%
\providecommand \citenamefont [1]{#1}%
\providecommand \href@noop [0]{\@secondoftwo}%
\providecommand \href [0]{\begingroup \@sanitize@url \@href}%
\providecommand \@href[1]{\@@startlink{#1}\@@href}%
\providecommand \@@href[1]{\endgroup#1\@@endlink}%
\providecommand \@sanitize@url [0]{\catcode `\\12\catcode `\$12\catcode
  `\&12\catcode `\#12\catcode `\^12\catcode `\_12\catcode `\%12\relax}%
\providecommand \@@startlink[1]{}%
\providecommand \@@endlink[0]{}%
\providecommand \url  [0]{\begingroup\@sanitize@url \@url }%
\providecommand \@url [1]{\endgroup\@href {#1}{\urlprefix }}%
\providecommand \urlprefix  [0]{URL }%
\providecommand \Eprint [0]{\href }%
\providecommand \doibase [0]{http://dx.doi.org/}%
\providecommand \selectlanguage [0]{\@gobble}%
\providecommand \bibinfo  [0]{\@secondoftwo}%
\providecommand \bibfield  [0]{\@secondoftwo}%
\providecommand \translation [1]{[#1]}%
\providecommand \BibitemOpen [0]{}%
\providecommand \bibitemStop [0]{}%
\providecommand \bibitemNoStop [0]{.\EOS\space}%
\providecommand \EOS [0]{\spacefactor3000\relax}%
\providecommand \BibitemShut  [1]{\csname bibitem#1\endcsname}%
\let\auto@bib@innerbib\@empty
\bibitem [{\citenamefont {Parrott}\ \emph {et~al.}(2018)\citenamefont
  {Parrott}, \citenamefont {Green}, \citenamefont {Milot}, \citenamefont
  {Johnston}, \citenamefont {Snaith},\ and\ \citenamefont
  {Herz}}]{Parrott2018}%
  \BibitemOpen
  \bibfield  {author} {\bibinfo {author} {\bibfnamefont {E.~S.}\ \bibnamefont
  {Parrott}}, \bibinfo {author} {\bibfnamefont {T.}~\bibnamefont {Green}},
  \bibinfo {author} {\bibfnamefont {R.~L.}\ \bibnamefont {Milot}}, \bibinfo
  {author} {\bibfnamefont {M.~B.}\ \bibnamefont {Johnston}}, \bibinfo {author}
  {\bibfnamefont {H.~J.}\ \bibnamefont {Snaith}}, \ and\ \bibinfo {author}
  {\bibfnamefont {L.~M.}\ \bibnamefont {Herz}},\ }\href@noop {} {\bibfield
  {journal} {\bibinfo  {journal} {Adv. Funct. Mater.}\ }\textbf {\bibinfo
  {volume} {28}},\ \bibinfo {pages} {1802803} (\bibinfo {year}
  {2018})}\BibitemShut {NoStop}%
\bibitem [{\citenamefont {Alferov}\ \emph {et~al.}(1969)\citenamefont
  {Alferov}, \citenamefont {Portnoi},\ and\ \citenamefont
  {Rogachev}}]{Alferov1969}%
  \BibitemOpen
  \bibfield  {author} {\bibinfo {author} {\bibfnamefont {Z.~I.}\ \bibnamefont
  {Alferov}}, \bibinfo {author} {\bibfnamefont {E.~M.}\ \bibnamefont
  {Portnoi}}, \ and\ \bibinfo {author} {\bibfnamefont {A.~A.}\ \bibnamefont
  {Rogachev}},\ }\href@noop {} {\bibfield  {journal} {\bibinfo  {journal} {Sov.
  Phys. Semicond.}\ }\textbf {\bibinfo {volume} {2}},\ \bibinfo {pages} {1001}
  (\bibinfo {year} {1969})}\BibitemShut {NoStop}%
\bibitem [{\citenamefont {Baranovskii}\ and\ \citenamefont
  {Efros}(1978)}]{Baranovskii1978}%
  \BibitemOpen
  \bibfield  {author} {\bibinfo {author} {\bibfnamefont {S.~D.}\ \bibnamefont
  {Baranovskii}}\ and\ \bibinfo {author} {\bibfnamefont {A.~L.}\ \bibnamefont
  {Efros}},\ }\href@noop {} {\bibfield  {journal} {\bibinfo  {journal} {Sov.
  Phys. Semicond.}\ }\textbf {\bibinfo {volume} {12}},\ \bibinfo {pages} {1328}
  (\bibinfo {year} {1978})}\BibitemShut {NoStop}%
\bibitem [{\citenamefont {Raikh}\ \emph {et~al.}(1990)\citenamefont {Raikh},
  \citenamefont {Baranovskii},\ and\ \citenamefont {Shklovskii}}]{Raikh1990}%
  \BibitemOpen
  \bibfield  {author} {\bibinfo {author} {\bibfnamefont {M.~E.}\ \bibnamefont
  {Raikh}}, \bibinfo {author} {\bibfnamefont {S.~D.}\ \bibnamefont
  {Baranovskii}}, \ and\ \bibinfo {author} {\bibfnamefont {B.~I.}\ \bibnamefont
  {Shklovskii}},\ }\href {\doibase 10.1103/PhysRevB.41.7701} {\bibfield
  {journal} {\bibinfo  {journal} {Phys. Rev. B}\ }\textbf {\bibinfo {volume}
  {41}},\ \bibinfo {pages} {7701} (\bibinfo {year} {1990})}\BibitemShut
  {NoStop}%
\bibitem [{\citenamefont {Wiemer}\ \emph {et~al.}(2016)\citenamefont {Wiemer},
  \citenamefont {Jandieri}, \citenamefont {Koch}, \citenamefont {Gebhard},\
  and\ \citenamefont {Baranovskii}}]{Wiemer2016}%
  \BibitemOpen
  \bibfield  {author} {\bibinfo {author} {\bibfnamefont {M.}~\bibnamefont
  {Wiemer}}, \bibinfo {author} {\bibfnamefont {K.}~\bibnamefont {Jandieri}},
  \bibinfo {author} {\bibfnamefont {M.}~\bibnamefont {Koch}}, \bibinfo {author}
  {\bibfnamefont {F.}~\bibnamefont {Gebhard}}, \ and\ \bibinfo {author}
  {\bibfnamefont {S.~D.}\ \bibnamefont {Baranovskii}},\ }\href@noop {}
  {\bibfield  {journal} {\bibinfo  {journal} {Phys. Status Solidi RRL}\
  }\textbf {\bibinfo {volume} {10}},\ \bibinfo {pages} {911} (\bibinfo {year}
  {2016})}\BibitemShut {NoStop}%
\bibitem [{\citenamefont {Masenda}\ \emph {et~al.}(2021)\citenamefont
  {Masenda}, \citenamefont {Schneider}, \citenamefont {Aly}, \citenamefont
  {Machchhar}, \citenamefont {Usman}, \citenamefont {Meerholz}, \citenamefont
  {Gebhard}, \citenamefont {Baranovskii},\ and\ \citenamefont
  {Koch}}]{Masenda2021}%
  \BibitemOpen
  \bibfield  {author} {\bibinfo {author} {\bibfnamefont {H.}~\bibnamefont
  {Masenda}}, \bibinfo {author} {\bibfnamefont {L.~M.}\ \bibnamefont
  {Schneider}}, \bibinfo {author} {\bibfnamefont {M.~A.}\ \bibnamefont {Aly}},
  \bibinfo {author} {\bibfnamefont {S.~J.}\ \bibnamefont {Machchhar}}, \bibinfo
  {author} {\bibfnamefont {A.}~\bibnamefont {Usman}}, \bibinfo {author}
  {\bibfnamefont {K.}~\bibnamefont {Meerholz}}, \bibinfo {author}
  {\bibfnamefont {F.}~\bibnamefont {Gebhard}}, \bibinfo {author} {\bibfnamefont
  {S.~D.}\ \bibnamefont {Baranovskii}}, \ and\ \bibinfo {author} {\bibfnamefont
  {M.}~\bibnamefont {Koch}},\ }\href@noop {} {\bibfield  {journal} {\bibinfo
  {journal} {Adv. Electron. Mater.}\ ,\ \bibinfo {pages} {2100196}} (\bibinfo
  {year} {2021})}\BibitemShut {NoStop}%
\bibitem [{\citenamefont {Weber}\ \emph {et~al.}(2016)\citenamefont {Weber},
  \citenamefont {Charles},\ and\ \citenamefont {Weller}}]{Weber2016}%
  \BibitemOpen
  \bibfield  {author} {\bibinfo {author} {\bibfnamefont {O.~J.}\ \bibnamefont
  {Weber}}, \bibinfo {author} {\bibfnamefont {B.}~\bibnamefont {Charles}}, \
  and\ \bibinfo {author} {\bibfnamefont {M.~T.}\ \bibnamefont {Weller}},\
  }\href {\doibase 10.1039/C6TA06607K} {\bibfield  {journal} {\bibinfo
  {journal} {J. Mater. Chem. A}\ }\textbf {\bibinfo {volume} {4}},\ \bibinfo
  {pages} {15375} (\bibinfo {year} {2016})}\BibitemShut {NoStop}%
\bibitem [{\citenamefont {Zong}\ \emph {et~al.}(2017)\citenamefont {Zong},
  \citenamefont {Wang}, \citenamefont {Zhang}, \citenamefont {Ju},
  \citenamefont {Zeng}, \citenamefont {Sun}, \citenamefont {Zhou},\ and\
  \citenamefont {Padture}}]{Zong2017}%
  \BibitemOpen
  \bibfield  {author} {\bibinfo {author} {\bibfnamefont {Y.}~\bibnamefont
  {Zong}}, \bibinfo {author} {\bibfnamefont {N.}~\bibnamefont {Wang}}, \bibinfo
  {author} {\bibfnamefont {L.}~\bibnamefont {Zhang}}, \bibinfo {author}
  {\bibfnamefont {M.-G.}\ \bibnamefont {Ju}}, \bibinfo {author} {\bibfnamefont
  {X.~C.}\ \bibnamefont {Zeng}}, \bibinfo {author} {\bibfnamefont {X.~W.}\
  \bibnamefont {Sun}}, \bibinfo {author} {\bibfnamefont {Y.}~\bibnamefont
  {Zhou}}, \ and\ \bibinfo {author} {\bibfnamefont {N.~P.}\ \bibnamefont
  {Padture}},\ }\href@noop {} {\bibfield  {journal} {\bibinfo  {journal}
  {Angew.\ Chem.}\ }\textbf {\bibinfo {volume} {129}},\ \bibinfo {pages} {1832}
  (\bibinfo {year} {2017})}\BibitemShut {NoStop}%
\bibitem [{\citenamefont {Muhammad}\ \emph {et~al.}(2020)\citenamefont
  {Muhammad}, \citenamefont {Liu}, \citenamefont {Ahmad}, \citenamefont
  {Asadabadi}, \citenamefont {Franchini},\ and\ \citenamefont
  {Ahmad}}]{Muhammad2020}%
  \BibitemOpen
  \bibfield  {author} {\bibinfo {author} {\bibfnamefont {Z.}~\bibnamefont
  {Muhammad}}, \bibinfo {author} {\bibfnamefont {P.}~\bibnamefont {Liu}},
  \bibinfo {author} {\bibfnamefont {R.}~\bibnamefont {Ahmad}}, \bibinfo
  {author} {\bibfnamefont {S.~J.}\ \bibnamefont {Asadabadi}}, \bibinfo {author}
  {\bibfnamefont {C.}~\bibnamefont {Franchini}}, \ and\ \bibinfo {author}
  {\bibfnamefont {I.}~\bibnamefont {Ahmad}},\ }\href@noop {} {\bibfield
  {journal} {\bibinfo  {journal} {Phys. Chem. Chem. Phys.}\ }\textbf {\bibinfo
  {volume} {22}},\ \bibinfo {pages} {11943} (\bibinfo {year}
  {2020})}\BibitemShut {NoStop}%
\bibitem [{\citenamefont {Wang}\ \emph {et~al.}(2020)\citenamefont {Wang},
  \citenamefont {Xiao},\ and\ \citenamefont {Wang}}]{Wang2020}%
  \BibitemOpen
  \bibfield  {author} {\bibinfo {author} {\bibfnamefont {S.}~\bibnamefont
  {Wang}}, \bibinfo {author} {\bibfnamefont {W.}~\bibnamefont {Xiao}}, \ and\
  \bibinfo {author} {\bibfnamefont {F.}~\bibnamefont {Wang}},\ }\href@noop {}
  {\bibfield  {journal} {\bibinfo  {journal} {RSC Adv.,}\ }\textbf {\bibinfo
  {volume} {10}},\ \bibinfo {pages} {32364} (\bibinfo {year}
  {2020})}\BibitemShut {NoStop}%
\bibitem [{\citenamefont {Dal~Don}\ \emph {et~al.}(2004)\citenamefont
  {Dal~Don}, \citenamefont {Kohary}, \citenamefont {Tsitsishvili},
  \citenamefont {Kalt}, \citenamefont {Baranovskii},\ and\ \citenamefont
  {Thomas}}]{DalDon2004}%
  \BibitemOpen
  \bibfield  {author} {\bibinfo {author} {\bibfnamefont {B.}~\bibnamefont
  {Dal~Don}}, \bibinfo {author} {\bibfnamefont {K.}~\bibnamefont {Kohary}},
  \bibinfo {author} {\bibfnamefont {E.}~\bibnamefont {Tsitsishvili}}, \bibinfo
  {author} {\bibfnamefont {H.}~\bibnamefont {Kalt}}, \bibinfo {author}
  {\bibfnamefont {S.~D.}\ \bibnamefont {Baranovskii}}, \ and\ \bibinfo {author}
  {\bibfnamefont {P.}~\bibnamefont {Thomas}},\ }\href {\doibase
  10.1103/PhysRevB.69.045318} {\bibfield  {journal} {\bibinfo  {journal} {Phys.
  Rev. B}\ }\textbf {\bibinfo {volume} {69}},\ \bibinfo {pages} {045318}
  (\bibinfo {year} {2004})}\BibitemShut {NoStop}%
\bibitem [{\citenamefont {Wright}\ \emph {et~al.}(2017)\citenamefont {Wright},
  \citenamefont {Milot}, \citenamefont {Eperon}, \citenamefont {Snaith},
  \citenamefont {Johnston},\ and\ \citenamefont {Herz}}]{Wright2017}%
  \BibitemOpen
  \bibfield  {author} {\bibinfo {author} {\bibfnamefont {A.~D.}\ \bibnamefont
  {Wright}}, \bibinfo {author} {\bibfnamefont {R.~L.}\ \bibnamefont {Milot}},
  \bibinfo {author} {\bibfnamefont {G.~E.}\ \bibnamefont {Eperon}}, \bibinfo
  {author} {\bibfnamefont {H.~J.}\ \bibnamefont {Snaith}}, \bibinfo {author}
  {\bibfnamefont {M.~B.}\ \bibnamefont {Johnston}}, \ and\ \bibinfo {author}
  {\bibfnamefont {L.~M.}\ \bibnamefont {Herz}},\ }\href@noop {} {\bibfield
  {journal} {\bibinfo  {journal} {Adv. Funct. Mater.}\ }\textbf {\bibinfo
  {volume} {27}},\ \bibinfo {pages} {1700860} (\bibinfo {year}
  {2017})}\BibitemShut {NoStop}%
\end{thebibliography}

%

\end{document}